\documentclass[twocolumn,nofootinbib]{revtex4}

\usepackage{epsfig}
\usepackage{amsmath,amsfonts,amssymb}

\providecommand{\openone}{\leavevmode\hbox{\small1\kern-3.8pt\normalsize1}}

\newcommand{\esx}{\langle S_1 \rangle}
\newcommand{\esy}{\langle S_2 \rangle}
\newcommand{\esxy}{\langle S_{1,2} \rangle}
\newcommand{\esi}{\langle S_k \rangle}
\newcommand{\esz}{\langle S_3 \rangle}
\newcommand{\etz}{\langle T_0 \rangle}
\newcommand{\eax}{\langle A_1 \rangle}
\newcommand{\eay}{\langle A_2 \rangle}
\newcommand{\eaxy}{\langle A_{1,2} \rangle}
\newcommand{\eai}{\langle A_k \rangle}
\newcommand{\ebx}{\langle B_1 \rangle}
\newcommand{\eby}{\langle B_2 \rangle}
\newcommand{\ebi}{\langle B_k \rangle}

\newcommand{\oh}{\textstyle \frac{1}{2}}
\newcommand{\ot}{\textstyle \frac{1}{3}}

\newcommand{\otst}{\textstyle \frac{1}{2 \sqrt 2}}
\newcommand{\ost}{\textstyle \frac{1}{\sqrt 2}}
\newcommand{\oss}{\textstyle \frac{1}{\sqrt 6}}

\renewcommand{\Re}{\operatorname{Re}}
\renewcommand{\Im}{\operatorname{Im}}
\newcommand{\sign}{\operatorname{sign}}

\begin{document}

\title{Breaking down the entire $W$ boson spin observables from its decay}

\author{J. A. Aguilar--Saavedra}
\affiliation{Departamento de F\'{\i}sica Te\'orica y del Cosmos, Universidad de Granada, E-18071 Granada, Spain}

\author{J. Bernab\'eu} 
\affiliation{Departament de F\'{\i}sica Te\'orica, Universitat de Val\'encia, 
E-46100, Burjassot, Spain}
\affiliation{IFIC, Universitat de Val\'encia-CSIC, E-46980, Paterna, Spain}

\begin{abstract}
We discuss the eight independent spin observables for the $W$ boson in terms of its vector and tensor polarizations and identify the angular distributions and asymmetries able to separate them in collider experiments. The results are applied to the study of polarized top quark decays and diboson resonances. These novel observables are of great value for disentangling new physics mechanisms in $W$ boson production.
\end{abstract}

\maketitle

The discovery by the ATLAS~\cite{Aad:2012tfa} and CMS~\cite{Chatrchyan:2012xdj} experiments at the Large Hadron Collider (LHC) of a new particle with a mass of 125 GeV and the expected properties of the Higgs boson has marked the programme of high energy physics for the next coming years. On one side, it is mandatory to be precise enough in the measurement of its properties in order to ascertain its nature 
as remnant of the electroweak symmetry breaking mechanism. On the other hand, the search of new physics beyond the standard model (SM), related or not to this symmetry breaking problem, makes the $W$ boson ---as mediator of the weak charged interactions---
a privileged particle to be a messenger of the scenario chosen by Nature. In addition, new unknown production mechanisms will leave their imprint in its detailed properties.

Interestingly enough, being a spin-1 particle, the $W$ boson provides 8 additional observables, beyond
the number of events, at our disposal to probe new physics and even discriminate among new physics models. As a consequence, it is of the highest importance to give an unambiguous procedure for extracting these spin properties from the experiments. This is the main motivation of this work. Some partial treatments have appeared in the literature~\cite{AguilarSaavedra:2010nx}, and also for specific reactions, like $e^+ e^-$ annihilation~\cite{Gounaris:1992kp}, but a direct methodology for connecting the decay distributions to the three polarizations and five alignments ---or tensor polarizations--- of the $W$ boson is lacking. Our results will be first applied to the decay of the top quark, which is the heaviest known particle and thus widely regarded as a promising probe of new physics~\cite{Bernreuther:2008ju,Han:2008xb,Aguilar-Saavedra:2014kpa}. Subsequently, we will consider the production of a $WZ/WW$ diboson resonance, motivated by a recent $3.4\sigma$ excess reported by the ATLAS Collaboration~\cite{Aad:2015owa}.

The role of the various mechanisms for the production of the $W$ boson is to describe its spin state in the form of a $3 \times 3$ density matrix, Hermitian with unit trace and positive semidefinite. By fixing a coordinate system $(x, y, z)$ in its rest reference frame, it can be written in terms of the three spin operator components $S_M$ and the five tensor operator components $T_M$ as
\begin{equation}
\rho = {\ot} \openone + \oh \displaystyle \sum_{M=-1}^1 \langle S_M \rangle^* S_M + \displaystyle \sum_{M=-2}^2  \langle T_{M} \rangle^* T_{M} \,,
\label{ec:rhoW}
\end{equation}
with $S_{\pm 1} = \mp \ost (S_1 \pm i S_2)$, $S_0 = S_3$ the spin operators in the spherical basis and $T_M$ five rank 2 irreducible tensors, which are built from $S_M$ as
\begin{align}
& T_{\pm 2} = S_{\pm1}^2 \,, \quad T_{\pm 1} = \ost \left[ S_{\pm 1} S_0 + S_0 S_{\pm 1} \right] \,, \notag \\
& T_{0} = \oss \left[ S_{+1} S_{-1} + S_{-1} S_{+1} + 2 S_0^2 \right] \,.
\end{align} 
Their expectation values $\langle S_M \rangle$ and $\langle T_M \rangle$ are the multipole parameters corresponding to the three polarization and five alignment components.
More conveniently, the second term in Eq.~(\ref{ec:rhoW}) can be rewritten using the spin operators in the Cartesian basis, and the third one defining the Hermitian operators
\begin{align}
A_1 = \oh (T_1-T_{-1}) \,, && A_2 = {\textstyle \frac{1}{2 i}} (T_1 + T_{-1}) \,, \notag \\
B_1 = \oh (T_2+T_{-2}) \,, && B_2 = {\textstyle \frac{1}{2 i}} (T_2 - T_{-2}) \,.
\end{align}
Explicitly, the $W$ boson density matrix parametrized in terms of expected values of observables reads
\begin{align}
& \rho_{\pm 1 \pm 1} = \ot \pm \oh \esz + \oss \etz \,,  \notag \\
& \rho_{\pm 10} = \otst \left[ \esx \mp i \esy \right] \mp \ost \left[ \eax \mp i \eay \right]  \,, \notag \\
& \rho_{00} = \ot - \textstyle \frac{2}{\sqrt 6} \etz \,, \quad \rho_{1\, -1} = \ebx - i \eby \,,
\label{ec:rhoexp}
\end{align}
and $\rho_{m'm}=\rho_{mm'}^*$.
This density matrix entirely determines the angular distribution of the $W$ decay products in $W^\pm \to \ell^\pm \nu$, with $\ell=e,\mu$. Using the helicity formalism of Jacob and Wick~\cite{Jacob:1959at}, the amplitude for the decay of a $W$ boson with third spin component $m$ into charged leptons $\ell$, $\nu$ with respective helicities $\lambda_1$, $\lambda_2$ is written as
\begin{equation}
\mathcal{M}_{m \lambda_1 \lambda_2} = a_{\lambda_1 \lambda_2} D_{m \lambda}^{1*} (\phi^*,\theta^*,0) \,,
\end{equation}
with $(\theta^*,\phi^*)$ the polar and azimuthal angles of the charged lepton momentum in the $W$ boson rest frame, $\lambda = \lambda_1 - \lambda_2$ and $D_{m'm}^j (\alpha,\beta,\gamma) =  e^{-i \alpha m'} e^{-i \gamma m} d_{m'm}^j (\beta)$ the so-called Wigner $D$ functions~\cite{wigner};  $a_{\lambda_1 \lambda_2}$ are constants, and all the dependence of the amplitude on angular variables of the final state products is given by the $D$ functions.
For the SM left-handed interaction of the $W$ boson with the charged leptons, and taking them massless, one has $(\lambda_1,\lambda_2) = (\pm 1/2, \mp 1/2)$ for $W^\pm \to \ell^\pm \nu$. Hence, the fully differential $W$ decay width can be written as
\begin{equation}
\frac{d\Gamma}{d\!\cos\theta^* d\phi^*}  = C \sum_{m,m'} \rho_{mm'} e^{i  (m-m')\phi^*} d_{m\lambda}^{1} (\theta^*) d_{m'\lambda}^{1}(\theta^*) 
\label{ec:dist2d}
\end{equation}
with $C$ a common prefactor including the phase space and the single non-zero $a_{\lambda_1 \lambda_2}$. Here, $\lambda = 1$ $ (-1)$ for $W^+$ ($W^-$) decays and the functions $d_{m'm}^j(\beta)$ are real, and so is the right hand side of Eq.~(\ref{ec:dist2d}). Explicitly, the normalised distribution reads
\begin{align}
& \frac{1}{\Gamma} \frac{d\Gamma}{d\!\cos\theta^* d\phi^*} = \frac{3}{8\pi} \left\{ 
\frac{1}{2} (1+\cos^2 \theta^*) + \esz \cos \theta^*
\right. \notag \\
& ~ + \left[ \frac{1}{6} - \frac{1}{\sqrt 6} \etz \right] \left( 1-3\cos^2 \theta^* \right) \notag \\
& ~ + \esx \cos \phi^* \sin \theta^* + \esy \sin \phi^* \sin \theta^* \notag \\ 
& ~ - \eax \cos \phi^* \sin 2\theta^* - \eay \sin \phi^* \sin 2\theta^* \notag \\
& \left. ~ + \ebx \cos 2 \phi^* \sin^2 \theta^* + \eby \sin 2 \phi^* \sin^2 \theta^* \right\} \,.
\label{ec:distfull}
\end{align} 
An analogous expansion has been used by the CDF~\cite{Aaltonen:2011nr} and CMS~\cite{Khachatryan:2015paa} Collaborations to study the dilepton angular distribution in $Z$ production, but an identification of the coefficients of the different terms with spin observables has not been provided --- indeed for the $Z$ boson the relation is more contrived since both dilepton helicity combinations $(\lambda_1,\lambda_2) = (\pm 1/2, \mp 1/2)$ contribute to the decay. This expansion is also implicit in~\cite{Gounaris:1992kp} but the correspondence with $W$ boson spin observables is not made (instead, predictions for some asymmetries are obtained for the specific case of $e^+ e^- \to W^+ W^-$ with general couplings). Polar asymmetries have been proposed in~\cite{AguilarSaavedra:2010nx}, to measure $\esx$ and $\esy$. Azimuthal asymmetries have also been proposed to test the spin of the decaying particles~\cite{Buckley:2007th,Boudjema:2009fz}, in particular the spin-1 nature of $W$ and $Z$ bosons can be tested by measuring the $\sin 2 \phi^*$ dependence~\cite{Buckley:2008pp}. (Within our framework, this amounts to measuring non-zero $\ebx$ or $\eby$, which clearly shows that the spin is larger or equal than one, given the rank of the irreducible tensor $T_2$.) But, as we have emphasized, a general method for connecting all $W$ boson polarization observables with angular distributions and asymemtries has not yet been provided.

The distribution (\ref{ec:distfull}) can in principle be used to determine all the $W$ boson spin observables with an eight-parameter fit. However, it is much simpler to project on certain directions to obtain one-dimensional distributions that depend on fewer parameters. This also allows for a cleaner extraction from experimental data without any loss of information. We do this in detail for the decay of a $W^+$ boson, and point the differences for $W^-$ boson decays. The asymmetries obtained here could as well be applied to the determination of all the parameters in the $Z$ boson distribution, extending current analyses~\cite{Khachatryan:2015paa} to the full set of eight coefficients.

\begin{figure}[htb]
\begin{center}
\begin{tabular}{ccc}
\includegraphics[height=3.9cm,clip=]{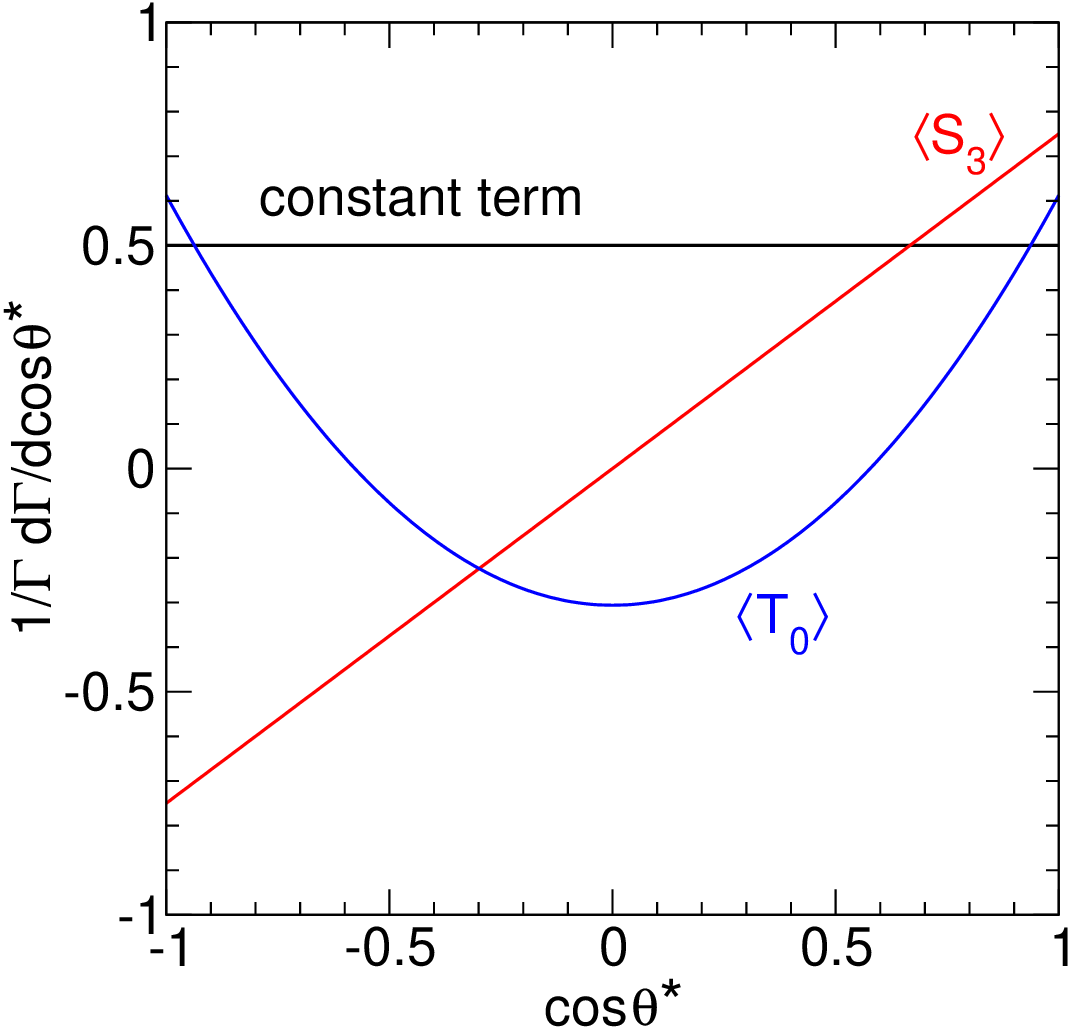} & &
\includegraphics[height=3.9cm,clip=]{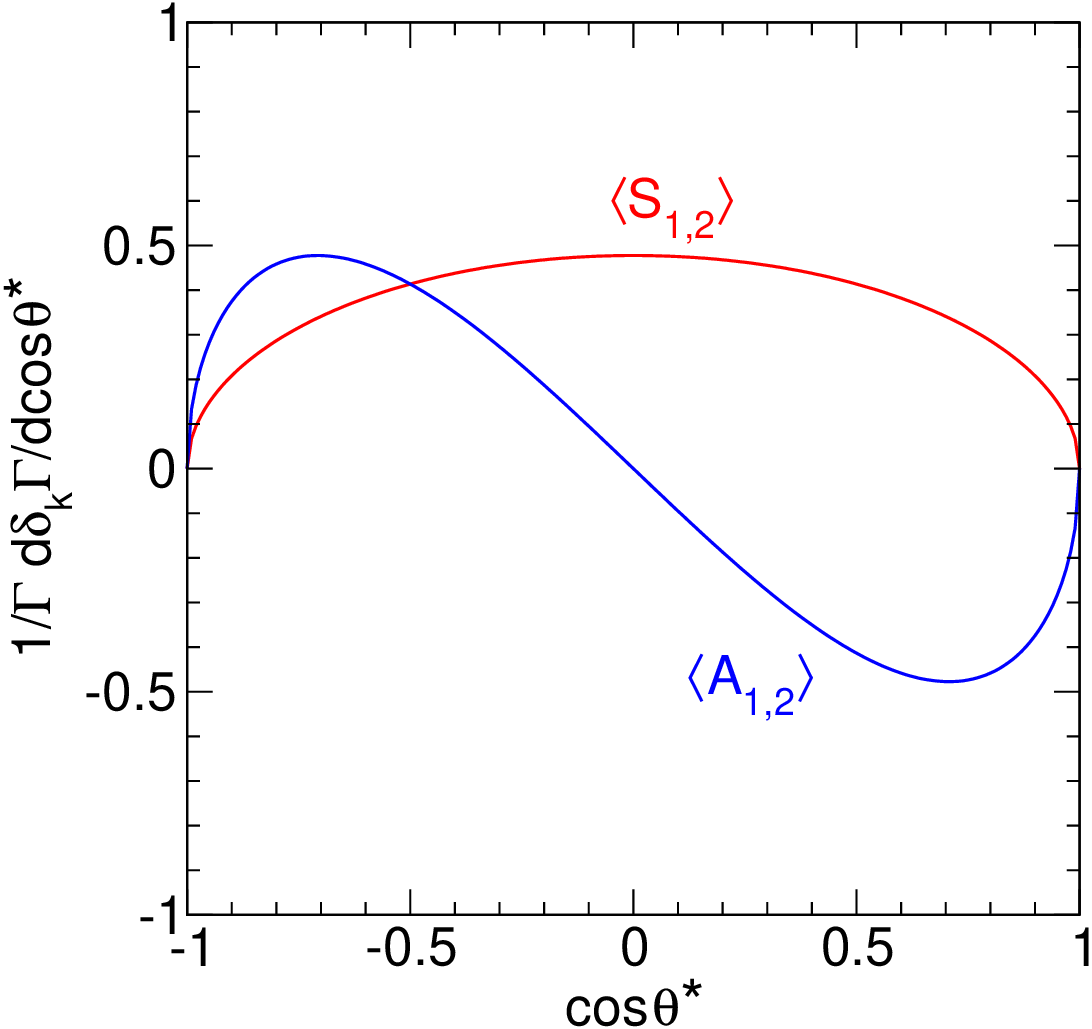} 
\end{tabular}
\caption{Left: functions multiplying $\esz$ and $\etz$ in the distribution~(\ref{ec:S3T0}), and constant term. Right: functions multiplying $\esxy$ and $\eaxy$ in the distributions~(\ref{ec:S1A1dist}). }
\label{fig:temp}
\end{center}
\end{figure}

Integrating over the azimuthal angle, the off-diagonal contributions vanish and the polar angle distribution reads
\begin{eqnarray}
\frac{1}{\Gamma} \frac{d\Gamma}{d\!\cos\theta^*} & = & \frac{3}{8} F_+ (1+\cos \theta^*)^2 + \frac{3}{4} F_0 \sin^2 \theta^* \notag \\
& & +  \frac{3}{8} F_- (1-\cos \theta^*)^2 \,,
\end{eqnarray}
where $F_+ = \rho_{11}$, $F_0 = \rho_{00}$ and $F_- = \rho_{-1-1}$ are the well-known $W$ polarisation fractions~\cite{Kane:1991bg,AguilarSaavedra:2010nx}, which are not independent but satisfy $F_+ + F_0 + F_- = 1$. In the common case that the $\hat z$ axis is taken in the direction of the $W$ momentum, $F_{\pm,0}$ are the helicity fractions.  These have been measured by several experiments for the $W$ bosons produced in top quark decays~\cite{Abazov:2010jn,Aad:2012ky,Aaltonen:2012lua,Chatrchyan:2013jna,Khachatryan:2014vma}. But instead of measuring three quantities that are not independent, it would be simpler to just use two independent ones: the longitudinal polarization $\esz$ and the alignment $\etz$, which obviously contain the very same information. Writing the explicit values of $\rho_{11}$, $\rho_{00}$ and $\rho_{-1-1}$ from Eqs.~(\ref{ec:rhoexp}), the resulting distribution is
\begin{equation}
\frac{1}{\Gamma} \frac{d\Gamma}{d\!\cos\theta^*} = \frac{1}{2} + \frac{3}{4} \esz \cos \theta^* + \frac{3}{4\sqrt 6} \etz \left[ 3 \cos^2 \theta^* - 1 \right] \,.
\label{ec:S3T0}
\end{equation}
With a two-parameter fit to the $\cos \theta^*$ distribution, $\esz$ and $\etz$ can be measured as well. For illustration, the functions multiplying these observables are shown in Fig.~\ref{fig:temp} (left). Alternatively, the two observables may be determined from asymmetries in the $\cos \theta^*$ distribution: the forward-backward (FB) asymmetry $A_\text{FB}$~\cite{Lampe:1995xb,delAguila:2002nf} and an ``edge-central'' asymmetry $A_\text{EC}$,
\begin{eqnarray}
A_\text{FB} & = & \frac{1}{\Gamma} \left[ \Gamma(\cos \theta^* > 0) - \Gamma(\cos \theta^* < 0) \right] = \frac{3}{4} \esz \,, \notag \\
A_\text{EC} & = & \frac{1}{\Gamma} \left[ \Gamma(|\cos \theta^*| > \oh ) - \Gamma(|\cos \theta^*| < \oh ) \right] \notag \\
& = & \frac{3}{8}  \sqrt{\frac{3}{2}}  \etz \,.
\label{ec:afbafc}
\end{eqnarray}
Whether this simpler approach also gives more precise results from the experimental point of view, once the experimental uncertainties are taken into account, has to be investigated with dedicated measurements.

In order to measure off-diagonal density matrix elements one must replace the integration over $\phi^*$ by an integration using a measure $f(\phi^*)$ that avoids the cancellation in the total integral, while selecting the desired observables. We first use $f_1(\phi^*) = \sign \cos \phi^*$, $f_2(\phi^*) = \sign \sin \phi^*$.
Experimentally, integrating with $f_1$ amounts to introducing a weight of $+1$ for events with $\phi^* \in \; ]-\pi/2,\pi/2[$, and of $-1$ for events with $\phi^* \in \; ]\pi/2,3\pi/2[$. Integrating with $f_2$ amounts to introducing a weight of $+1$ for events with $\phi^* \in \; ]0,\pi[$, and of $-1$ for events with $\phi^* \in \; ]\pi,2\pi[$. The resulting quantities, which we denote as $\delta_1 \Gamma$, $\delta_2 \Gamma$, have an angular distribution
\begin{equation}
\frac{1}{\Gamma} \frac{d(\delta_k \Gamma)}{d\!\cos\theta^*} = \frac{3}{2\pi} \esi \sin \theta^* - \frac{3}{2 \pi} \eai \sin 2 \theta^* \,,
\label{ec:S1A1dist}
\end{equation}
with $k=1,2$. The functions multiplying $\esi$ and $\eai$ are shown in Fig.~\ref{fig:temp} (right). These four observables can be measured by fitting the above distributions or, in a simpler way, by either integrating over $\cos \theta^*$ or doing the forward-backward asymmetry in this angle. The total integrals on the polar angle $\theta^*$ give
\begin{eqnarray}
\frac{\delta_1 \Gamma}{\Gamma} & = & \frac{1}{\Gamma}  \left[ \Gamma(\cos \phi^* > 0) - \Gamma(\cos \phi^* < 0) \right] = \frac{3}{4} \esx \,, \notag \\
\frac{\delta_2 \Gamma}{\Gamma} & = & \frac{1}{\Gamma}  \left[ \Gamma(\sin \phi^* > 0) - \Gamma(\sin \phi^* < 0) \right] = \frac{3}{4} \esy \,. \notag \\
\label{ec:s1s2}
\end{eqnarray}
Notice that $\delta_1 \Gamma/\Gamma$ and $\delta_2 \Gamma/\Gamma$  are merely the FB asymmetries with respect to the $\hat x$ and $\hat y$ axes, respectively~\cite{AguilarSaavedra:2010nx}, so the relation with $\esx$ and $\esy$ is expected. 
The FB asymmetries $A_\text{FB}^k$ in the $\delta_k \Gamma$ distributions~(\ref{ec:S1A1dist}) yield the other two observables,
\begin{eqnarray}
A_\text{FB}^1 & = &  \frac{1}{\Gamma}  \left[ \Gamma(\cos \phi^* \cos \theta^* > 0) - \Gamma(\cos \phi^* \cos \theta^* < 0) \right] \notag \\
& = & - \frac{2}{\pi} \eax \,, \notag \\ 
A_\text{FB}^2 & = &  \frac{1}{\Gamma}  \left[ \Gamma(\sin \phi^* \cos \theta^* > 0) - \Gamma(\sin \phi^* \cos \theta^* < 0) \right] \notag \\
& = & - \frac{2}{\pi} \eay \,. 
\label{ec:a1a2}
\end{eqnarray}

Let us now consider the measurement of $\ebi$. Integrating the doubly differential distribution with
$g_1(\phi^*) = \sign \cos 2\phi^*$, $g_2(\phi^*) = \sign \sin 2  \phi^*$,
all terms in Eq.~(\ref{ec:dist2d}) vanish except those involving $\rho_{\pm 1\mp 1}$. Furthermore, $g_1$ selects the terms proportional to $\ebx$ and $g_2$ the ones proportional to $\eby$, so the integration over $\cos \theta^*$ can be performed to get azimuthal asymmetries
\begin{eqnarray}
A_\phi^1 & = & \frac{1}{\Gamma} \left[ \Gamma(\cos 2 \phi^* > 0 ) - \Gamma(\cos 2\phi^* < 0) \right] = \frac{ 2}{\pi} \ebx \,, \notag \\
A_\phi^2 & = & \frac{1}{\Gamma} \left[ \Gamma(\sin 2 \phi^* > 0) - \Gamma(\sin 2\phi^* < 0) \right] = \frac{2}{\pi} \eby \,. \notag \\
\label{ec:aphi}
\end{eqnarray}

For $W^-$ boson decays the differential distributions are obtained from the ones presented by replacing $\sin \theta^* \to - \sin \theta^*$, $\cos \theta^* \to - \cos \theta^*$. The relations between asymmetries and observables in Eqs.~(\ref{ec:afbafc}), (\ref{ec:s1s2}), (\ref{ec:a1a2}) and (\ref{ec:aphi}) are the same for $W^-$ decays, with extra minus signs for $A_\text{FB}$ and $\delta_k \Gamma/\Gamma$, i.e. those asymmetries proportional to the expected values of the spin operators.

{\em Application to polarized top quark decays.}
The density matrix elements for $W$ bosons resulting from the decay of a fully polarized top quark were given in Ref.~\cite{AguilarSaavedra:2010nx} for a general $Wtb$ vertex including left and right vector and tensor couplings, using the helicity basis. Setting our $\hat z$ axis in the direction of the $W$ boson momentum in the top quark rest frame, and the top spin direction in the $\hat x- \hat z$ plane,
$\vec s_t = (\sin \theta,0,\cos \theta)$, one has the (unnormalized) density matrix elements
\begin{align}
& \hat \rho_{\pm 1 \pm1} = \mathcal{B}_0 (1 \pm \cos \theta) \pm 2 \frac{q}{m_t} \mathcal{B}_1 (1 \pm \cos \theta) \,, \notag \\
& \hat \rho_{\pm 10} = \left[ \frac{m_t}{\sqrt 2 M_W} (\mathcal{C}_0 + i \mathcal{D}_0) \pm \frac{q}{\sqrt 2 M_W} (\mathcal{C}_1 + i \mathcal{D}_1) \right] \sin \theta \,, \notag \\
& \hat \rho_{00} = \mathcal{A}_0 + 2 \frac{q}{m_t} \mathcal{A}_1 \cos \theta \,, \quad  \hat \rho_{1-1}=0 \,, 
\end{align}
and $\hat \rho_{m'm}= \hat \rho_{mm'}^*$, with $m_t$ and $M_W$ the top quark and $W$ boson masses, respectively, $q$ the modulus of the $W$ boson 3-momentum in the top quark rest frame, and $\mathcal{A}_{0,1}$, $\mathcal{B}_{0,1}$, $\mathcal{C}_{0,1}$, $\mathcal{D}_{0,1}$ constants that in general depend on the $Wtb$ couplings. Their explicit expressions were given in Ref.~\cite{AguilarSaavedra:2010nx}. (For top anti-quark decays one has to reverse the sign of the  $\mathcal{B}_1$, $\mathcal{D}_0$, $\mathcal{C}_1$ and $\mathcal{A}_1$ terms.)
For simplicity we consider here the quantities integrated over $\theta$, despite an analysis dependent on this angle, or with $\theta$ restricted to some interval, is possible with enough statistics.
Integrating over $\cos \theta$ and normalizing to the $W \to \ell \nu$ width, one has
\begin{align}
& \esz = 4 \frac{q}{m_t} \frac{\mathcal{B}_1}{\mathcal{A}_0+2 \mathcal{B}_0} \,, \quad
 \etz = \sqrt \frac{2}{3} \frac{\mathcal{B}_0-\mathcal{A}_0}{\mathcal{A}_0+2\mathcal{B}_0}  \,, \notag \\
& \esx = \frac{\pi}{2} \frac{m_t}{M_W} \frac{\mathcal{C}_0}{\mathcal{A}_0+2 \mathcal{B}_0} \,, \quad
\esy = - \frac{\pi}{2} \frac{q}{M_W} \frac{\mathcal{D}_1}{\mathcal{A}_0+2 \mathcal{B}_0} \,, \notag \\
& \eax = - \frac{\pi}{4} \frac{q}{M_W} \frac{\mathcal{C}_1}{\mathcal{A}_0+2 \mathcal{B}_0} \,, \quad
\eay = \frac{\pi}{4} \frac{m_t}{M_W} \frac{\mathcal{D}_0}{\mathcal{A}_0+2\mathcal{B}_0} \,, \notag \\
\label{ec:obstop}
\end{align}
and $ \ebx = \eby = 0$ since $\hat \rho_{\pm 1 \mp 1} = 0$.

High energy collider processes do not produce fully polarized top quarks but with a partial polarization. For example, in $t$-channel single top quark production at the LHC, top quarks have a longitudinal polarization $P_z \simeq 0.9$ in the direction of the spectator jet. In this process, the top transverse and normal polarizations $P_{x,y}$ vanish~\cite{Aguilar-Saavedra:2014eqa}, so one can take into account in Eqs.~(\ref{ec:obstop}) the degree of top quark polarization by simply introducing a $P_z$ factor in the expressions of $\esxy$ and $\eaxy$.  We remark that $P_z$ can be measured in a model-independent fashion~\cite{AguilarSaavedra:2012xe}, so there is no need to assume that it takes the SM value in order to measure the $W$ boson spin observables. Four of these observables are nonzero in the SM, and their tree-level values for $P_z = 1$ are collected in Table~\ref{tab:pol1}.

\begin{table}[htb]
\begin{center}
\begin{tabular}{l|ccccccc}
\hline
\hline
& $\esx$ & $\esy$ & $\esz$ & $\etz$ & $\eax$ & $\eay$ \\
\hline
SM & 0.510 & 0 & -0.302 & -0.445 & 0.255 & 0  \\
$g_R = 0.03$ & 0.500 & 0 & -0.278 & -0.472 & 0.249 & 0  \\
$g_R = 0.10 \,i$ & 0.507 & -0.084 & -0.284 & -0.434 & 0.253 & -0.042  \\
\hline
\hline
\end{tabular}
\caption{SM tree-level predictions for $W$ spin observables in polarized top quark decays, and some predictions in the presence of an anomalous $Wtb$ dipole interaction.}
\label{tab:pol1}
\end{center}
\end{table}

The experimental values of $\esz$ and $\etz$ can be determined from current measurements by using their relation with the $W$ helicity fractions,
\begin{align}
& \esz = F_+ - F_-  \,, \quad \etz = \frac{1}{\sqrt 6} \left[ 1-3 F_0 \right] \,.
\end{align}
Using the  most precise measurements, obtained by the CMS Collaboration in $t \bar t$ production at 7 TeV~\cite{Chatrchyan:2013jna}, we obtain $\esz = -0.302 \pm 0.036$, $\etz = -0.427 \pm 0.054$, in good agreement with the SM prediction. (Next-to-next-to-leading order predictions for $F_{\pm,0}$ are available~\cite{Czarnecki:2010gb}, but the differences with the tree level results are below the $0.3\sigma$ level and for consistency with the rest of observables we use the tree-level predictions in order to calculate $\esz$ and $\etz$.) Notice that the uncertainty on $\esz$, obtained here by error propagation, would be around a factor of two smaller with a dedicated measurement.
$\esx$ and $\eax$, both nonzero in the SM, have not been measured so far. In this respect, we point out that the measurement of $\eax$ in single top quark production would be one of the very few, if not the first, measurement of the $M>0$ tensor polarization of an elementarty particle.\footnote{The OPAL~\cite{Abbiendi:2000ei} and Delphi~\cite{Abdallah:2008sf} Collaborations measured the $W$ density matrix elements at the Large Electron Positron collider, as a tool to search for anomalous triple boson couplings. The measurements of $\Re \rho_{1-1}$ and $\Im \rho_{1-1}$ could be intepreted as measurements of tensor polarizations but the uncertainties were so large that it is difficult to claim that a measurement was made.} Remarkably, $\esx$, $\eax$ are measurable with current LHC data at 8 TeV, using the same signal identification and background rejection strategies of current measurements~\cite{Khachatryan:2014vma,ATLAS:2013ula}, but considering different angular distributions. The expected uncertainties are around $0.03-0.05$, i.e. of the same order as those for $\esz$ and $\etz$. $\esy$ and $\eay$ are nonzero only if there are anomalous imaginary contributions to the $Wtb$ vertex; among them, only $\esy$ has been experimentally probed in $t$-channel single top quark production at 7 TeV by the ATLAS Collaboration~\cite{ATLAS:2013ula}. Assuming the SM value for the top quark polarization (since it has not been yet measured independently), the measurement of $\delta_2 \Gamma/\Gamma$ yields $\esy = -0.045 \pm 0.106$. Here the uncertainty is dominated by the statistical error, so a large improvement, by a factor of two or three, is foreseen with already collected 8 TeV data.

These polarization observables are also sensitive to new physics in the top quark  decay. For illustration, we show in Table~\ref{tab:pol1} the predictions in the presence of a $Wtb$ dipole interaction of the form
\begin{equation}
-\frac{g}{\sqrt 2 M_W} g_R \bar b_L \sigma^{\mu \nu} t_R \partial_\mu W_\nu^- \,,
\end{equation}
whose strength is driven by the anomalous coupling $g_R$. This amplitude is particularly interesting because it interferes with the standard model amplitude in the massless $b$ limit. We select two values for this parameter, real and purely imaginary, which are well within the $1\sigma$ limits given by current measurements~\cite{Chatrchyan:2013jna,ATLAS:2013ula}. As it is apparent, the measurement of these additional polarization observables will also improve the sensitivity to new physics in the top quark decay.

{\em Application to diboson resonances.}
For a heavy resonance $R$ decaying into $WZ$ or $WW$ the $W$ boson spin observables can also be measured using the method outlined here. We consider as example a heavy scalar or vector resonance produced in quark-antiquark annihilation, focusing for definiteness on $W^+$ boson decays. We set our $\hat z$ axis in the direction of the $W^+$ momentum in the centre of mass frame. The remaining two axes are fixed by taking the $\hat x$ axis in the production plane, and requiring that the momentum of the diboson pair in the laboratory frame (used as a proxy for the direction of the initial quark momentum) has a positive $x$ component.

We give in Table~\ref{tab:pol2} the spin observables for a resonance with a mass of 2 TeV, as the diboson resonance peak found by the ATLAS Collaboration~\cite{Aad:2015owa}, and a moderate width of 50 GeV.
\begin{table}[t!]
\begin{center}
\begin{tabular}{l|cccccccc}
\hline
\hline
Resonance & $\esx$ & $\esy$ & $\esz$ & $\etz$ & $\eax$ & $\eay$ & $\ebx$ & $\eby$ \\
\hline
$W Z$ $J=0$ & 0 & 0 & 0 & -0.815 & 0 & 0 & 0 & 0 \\
$W Z$ $J=1$  & 0.117 & 0 & 0 & -0.801 & 0 & 0 & 0.003 & 0 \\
$W W$ $J=0$ & 0 & 0 & 0 & -0.815 & 0 & 0 & 0 & 0 \\
$W W$ $J=1$ & 0.099 & 0 & 0 & -0.800 & 0 & 0 & 0.003 & 0 \\
\hline
\hline
\end{tabular}
\caption{Spin observables for $W^+$ bosons resulting from the decay of a heavy scalar or vector $WZ/WW$ resonance. Entries numerically smaller than $10^{-3}$ are indicated as zero.}
\label{tab:pol2}
\end{center}
\end{table}%
The numerical calculations are performed with the Monte Carlo generator {\sc Protos}~\cite{protos}. It is known that for both heavy scalar or vector diboson resonances the resulting $W/Z$ bosons have a big longitudinal alignment of almost 100\%, with $\esz = 0$ and $\etz \simeq -2/\sqrt 6 \simeq -0.816$, i.e. they are mostly produced with zero helicity. We observe in Table~\ref{tab:pol2} that the differences in $\etz$ between scalar and vector resonances are indeed tiny and most probably unmeasurable. However, $\esx$ vanishes for a scalar while is of the order of $0.1$ for a vector. This difference could be exploited to determine the spin of the resonance, provided it is detected in the $W\to \ell \nu$ decay channel, as it should be if the ATLAS excess indeed corresponds to a $WZ$ or $WW$ resonance.

To summarize, in this Letter we have set a framework for the study of {\it all} spin observables of the $W$ boson from its decay. We have given two examples of the usefulness of these observables, for $W$ bosons from the decay of polarized top quarks and for diboson resonances. The potential applications are numerous, as the formalism presented here is completely general. Because $W$ bosons are frequent decay products of (new) heavy particles, this formalism is expected to be of great value in the search of new physics effects.

\begin{acknowledgments}
We thank the CERN Theory Unit for the hospitality extended to us while performing this 
work. This research has been supported by MINECO Projects FPA 2013-47836-C3-2-P,
FPA 2011-23596, Junta de Andaluc\'{\i}a Project FQM-101, Generalitat Valenciana Project
GV PROMETEO II 2013-017 and Severo Ochoa Excellence Centre Project 
SEV 2014-0398.
\end{acknowledgments}

\end{document}